%% file: BinaryPaper2.tex
\documentclass[twocolumn]{aastex631}
\usepackage{amsmath}
\usepackage{multirow}
\usepackage{url}
\usepackage{color}

\usepackage{amssymb}
\usepackage{graphicx}
\usepackage{upgreek}
\usepackage{mathtools}

\input{Macros.tex}

\shorttitle{Opportunistic constraints from continuous gravitational waves for neutron stars in binaries}
\shortauthors{Singh and Papa}

\begin{document}

\title{Opportunistic search for continuous gravitational waves from compact objects in long-period binaries}

\correspondingauthor{Avneet Singh}
\email{avneet.singh@aei.mpg.de}

\correspondingauthor{Maria Alessandra Papa}
\email{maria.alessandra.papa@aei.mpg.de}

\author{Avneet Singh}
\affiliation{Max-Planck-Institut f\"{u}r Gravitationsphysik
  (Albert-Einstein-Institut), D-30167 Hannover, Germany}
\affiliation{Leibniz Universit\"at Hannover, 30167 Hannover, Germany}

\author{Maria Alessandra Papa}
\affiliation{Max-Planck-Institut f\"{u}r Gravitationsphysik
  (Albert-Einstein-Institut), D-30167 Hannover, Germany}
\affiliation{Leibniz Universit\"at Hannover, 30167 Hannover, Germany}
\affiliation{Department of Physics, University of Wisconsin, Milwaukee, WI 53201, USA}

\begin{abstract}
Most all-sky searches for continuous gravitational waves assume the source to be isolated. In this paper, we allow for an unknown companion object in a long-period orbit and opportunistically use previous results from an all-sky search for isolated sources to constrain the continuous gravitational wave amplitude over a large and unexplored range of binary orbital parameters without explicitly performing a dedicated search for binary systems. The resulting limits are significantly more constraining than any existing upper limit for unknown binary systems, albeit the latter apply to different orbital parameter ranges \avn{that are computationally much costlier to explore}.
\end{abstract}
%\maketitle


\nolinenumbers %for arXiv 
%\linenumbers
\section{Introduction}
\label{section:intro}
No search for continuous gravitational waves so far has produced clear evidence of a signal, including broad-band and all-sky surveys \citep{LIGOScientific:2022pjk,LIGOScientific:2019yhl,Steltner:2020hfd,Dergachev:2022lnt,LIGOScientific:2021o3a,O2FalconHF,O2FalconLF,O2FalconMF}.

The main signals targeted by these searches are fast rotating neutron stars with an equatorial deformation, an ellipticity $\varepsilon = {{(\vams{I}{xx}-\vams{I}{yy})}/\vams{I}{zz}}$, where $\mathrm{I}$ is the moment of inertia tensor and $\hat{\mathrm{z}}$ is aligned with the spin of the star. These searches are typically hierarchical, with a large number of candidates followed up through various stages of increasing sensitivity.

The most sensitive all-sky searches have targeted emissions from isolated objects. With many known millisecond pulsars in binary systems, it is reasonable to wonder whether a signal might have been dismissed as inconsistent with the assumed isolated model during the follow-up, because it in fact came from an object in a binary system.

This paper takes the first step in using results from a continuous wave isolated neutron star search, to investigate emissions from neutron stars with companions.

We identify orbital parameter ranges, for which a signal from a binary system would have appeared as a signal from an isolated object in the results of Stage 0 of {\EatH} search \citep{Steltner:2020hfd}. In this range, based on such results \citep{Steltner:2020hfd}, we constrain the intrinsic gravitational wave amplitude from neutron stars in binary systems.  

\begin{figure}[ht!]
\centering\includegraphics[width=85mm]{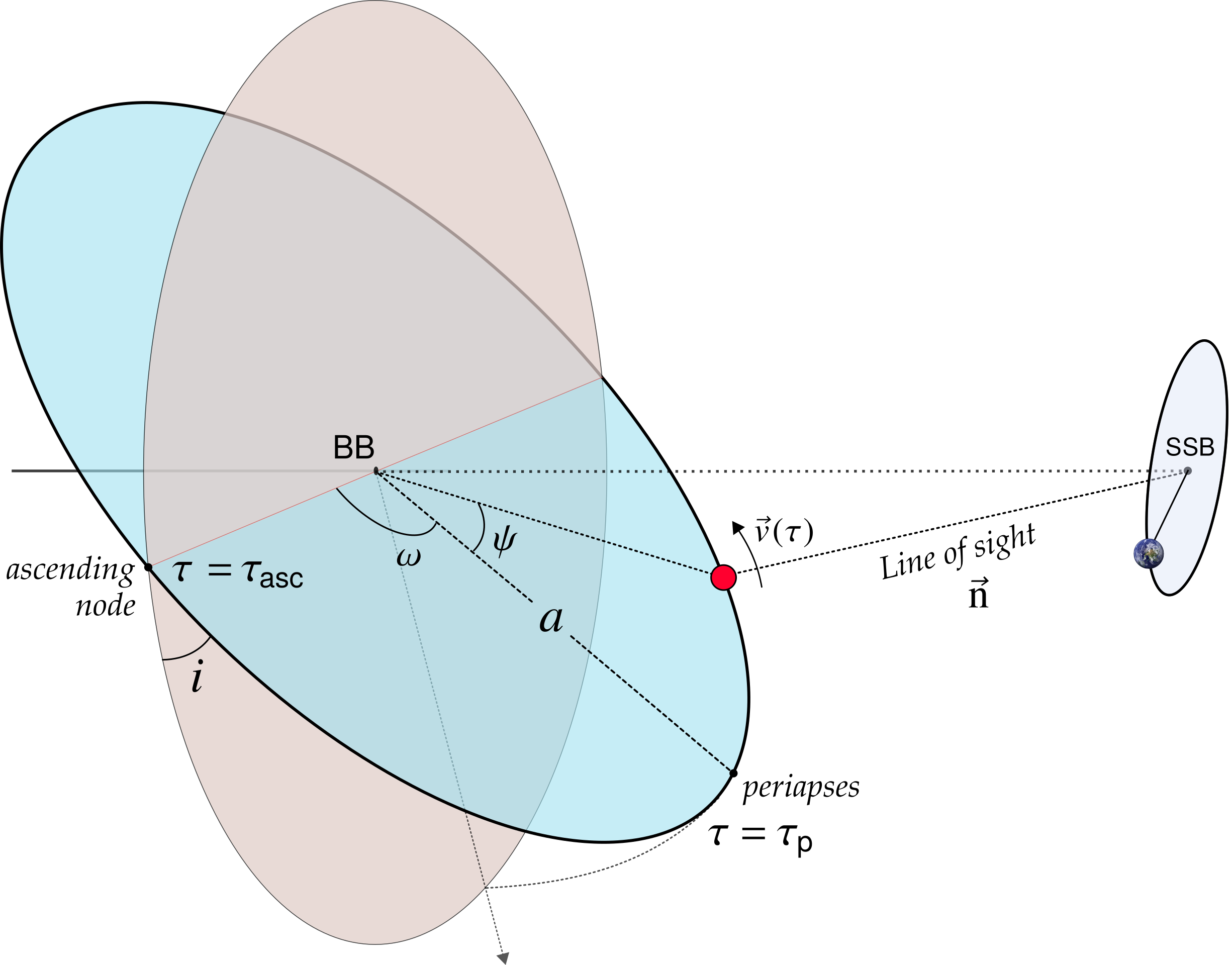}
\caption{{\,\small A nearly circular binary orbit. {{\it Abrrev}}: \textrm{BB} $\equiv$ {Binary Barycentre}, \textrm{SSB} $\equiv$ {Solar System Barycentre}. The projected semi-major axis is reported in units of time, lightseconds, i.e. $\vars{a}{p} = a\csin i/c$, and $\vams{P}{b}\sim 2\uppi a/|\vec{v}|$ is the period of the binary orbit.}}
\label{fig:orbit}
\end{figure}

\section{Exploring binary parameter space with isolated continuous wave searches}
\label{sec:searchresults} 
We perform a series of Monte-Carlo studies where continuous wave signals from sources in binary circular orbits are simulated in noiseless data. \avn{The simulated continuous wave signals are spread uniformly over the entire sky, in frequency between 20--585 Hz, in cosine of the neutron star orientation, $-1\leq\cos\iota\leq 1$, in polarization angle $|\psi|\leq \pi/4$, and log-uniformly in frequency derivative between $-2.6\times10^{-9}$ and $2.6\times10^{-10}$ Hz/s.
These ranges are defined by the {\EatH} search \citep{Steltner:2020hfd}.} We consider an exhaustive range of projected orbital semi-major axis, orbital period and time of ascending nodes: $\vars{a}{p}\in [10^{-2}, 10^6]$ lightseconds, $\vams{P}{b}\in [10^{-6},10^5]$ years and $\vars{\psi}{asc}= 0, {\uppi}/{6}, {\uppi}/{3}, {\uppi}/{2}$, with $\vars{\psi}{asc}=2\pi \vars{t}{asc}/\vams{P}{b}$. Figure \ref{fig:orbit} shows the various orbital parameters and clarifies the symbols. The values of $\vars{a}{p}$ and $\vams{P}{b}$ are log-uniformly distributed. 
 
\begin{figure}[ht!]
\centering\includegraphics[width=90mm]{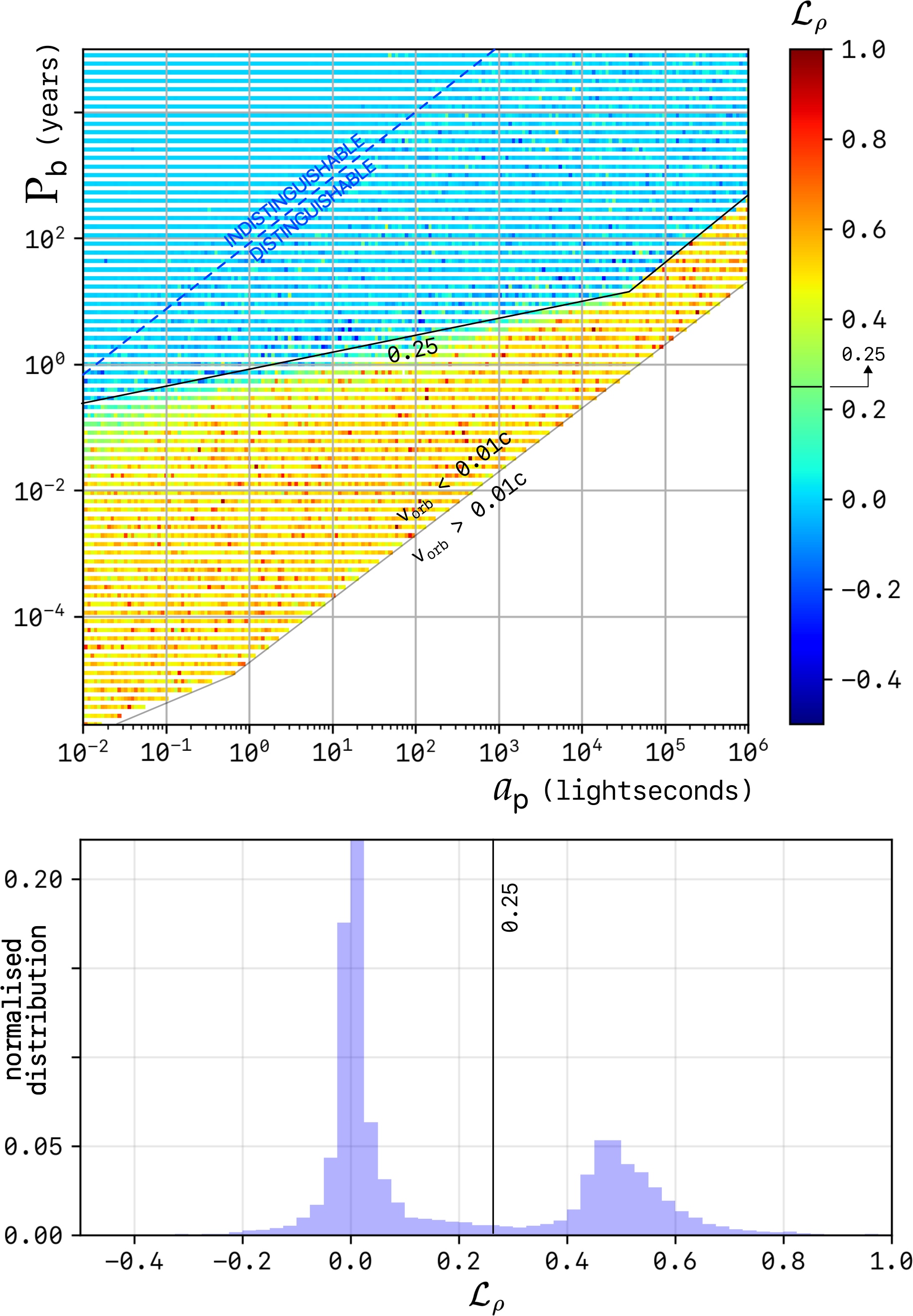}
\caption{Distributions of $\mathcal{L}_{\rho}$ of equation \eqref{eq:snrloss}, based on 15,000 injections and recoveries of simulated signals as described in the text with $\vars{\psi}{asc}= 0$; for different values of $\vars{\psi}{asc}$, the regions are very similar and the yellow higher-loss region is in fact slightly less extended.}
\label{fig:SNRchangeDistr}
\end{figure}

Each signal is searched for with the same semi-coherent $\TwoF$ search algorithm for isolated signals as used in Stage 0 \citep{Steltner:2020hfd}, yielding a value \avn{of {\bf{S}}ignal-to-{\bf{N}}oise {\bf{R}}atio (SNR) equal to $\rho_{\textrm{bin-iso}}$}, where the subscript means {\bf{bin}}ary signal searched with an {\bf{iso}}lated signal model. The same signal, but coming from an isolated object, is also simulated and searched for, and the $\rho_{\textrm{iso-iso}}$ is recorded. 
\avn{We define a quantity $\mathcal{L}_{\rho}$ encapsulating the {\SNR} change} for each signal searched as
\begin{equation}
\mathcal{L}_{\rho} = \frac{\rho^2_\mathrm{iso-iso} - \rho^2_\mathrm{bin-iso}}{\rho^2_\mathrm{max}},
\label{eq:snrloss}
\end{equation}
where $\rho_\mathrm{max}$ is the maximum possible {\SNR}, attainable for zero mismatch between signal and search template in all of the signal parameters. $\mathcal{L}_{\rho}$ can be positive or negative; the negative values occur when loss of SNR due to a mismatch in phase parameters is compensated by the modulation due to the orbital parameters. For every signal, we also record the distance between the signal's phase parameters and the phase parameters recovered by the search. Figure \ref{fig:SNRchangeDistr} shows the distributions of $\mathcal{L}_{\rho}$. 

We identify three regions in the ($\vars{a}{p}-\vams{P}{b}$) plane : {{\it a}}) {{\it indistinguishable}} region where $\mathcal{L}_{\rho} < 1\%$ {{\it b}}) {{\it distinguishable}} region where $\mathcal{L}_{\rho} \leq 25\%$ {{\it c}}) the remaining region where $\mathcal{L}_{\rho} \sim 100\%$. In the indistinguishable region, the location of the maximum detection statistic in the parameter space due to a binary signal is statistically at the same distance from the signal parameters as the maximum from an isolated signal; in the distinguishable region however, this is not the case. These regions are shown in Figure \ref{fig:SNRchangeDistr} and they are broadly consistent with our predictions in \citep{Binary1}, albeit the case studied here is more general \avn{due to the inclusion of randomly sampled $\vars{t}{asc}$ in binary orbital parameters. In addition, this study features a broader frequency range up to nearly 600 Hz as well as uniform sampling in the sky, whereas \citep{Binary1} extended only until 100 Hz in frequency and sampled only a small section of the sky}. The regions' boundaries are slightly different at the highest $\vars{a}{p}$ values, depending on the value of $\vars{\psi}{asc}$. We have picked the boundaries in the most conservative way, i.e. so that the conditions on $\mathcal{L}_{\rho}$ above hold for any value of $\vars{\psi}{asc}$, and in particular for $\vars{\psi}{asc}=0$.

Based on these findings, we use the results of Stage 0 of the Einstein@Home search for signals from isolated objects \citep{Steltner:2020hfd} to constrain the amplitude of continuous gravitational waves from neutron stars in binary systems in the distinguishable and indistinguishable regions in $\vars{a}{p}-\vams{P}{b}$ plane. We expect that the resulting upper limits will not be significantly less constraining than the limits set by \cite{Steltner:2020hfd}.  

In practice, we take the most significant candidate from the Stage 0 results of \cite{Steltner:2020hfd} in every 0.5 Hz band and find the value of the intrinsic gravitational wave amplitude $h_0^{90\%}$ such that 90\% of the population of signals with binary parameters in distinguishable and indistinguishable regions would have produced a detection statistic equal to the observed value. This is the standard definition of the 90\% confidence gravitational wave amplitude upper limit used in many continuous waves searches, including \citep{Steltner:2020hfd}. 

\begin{figure}[ht!]
\centering\includegraphics[width=85mm]{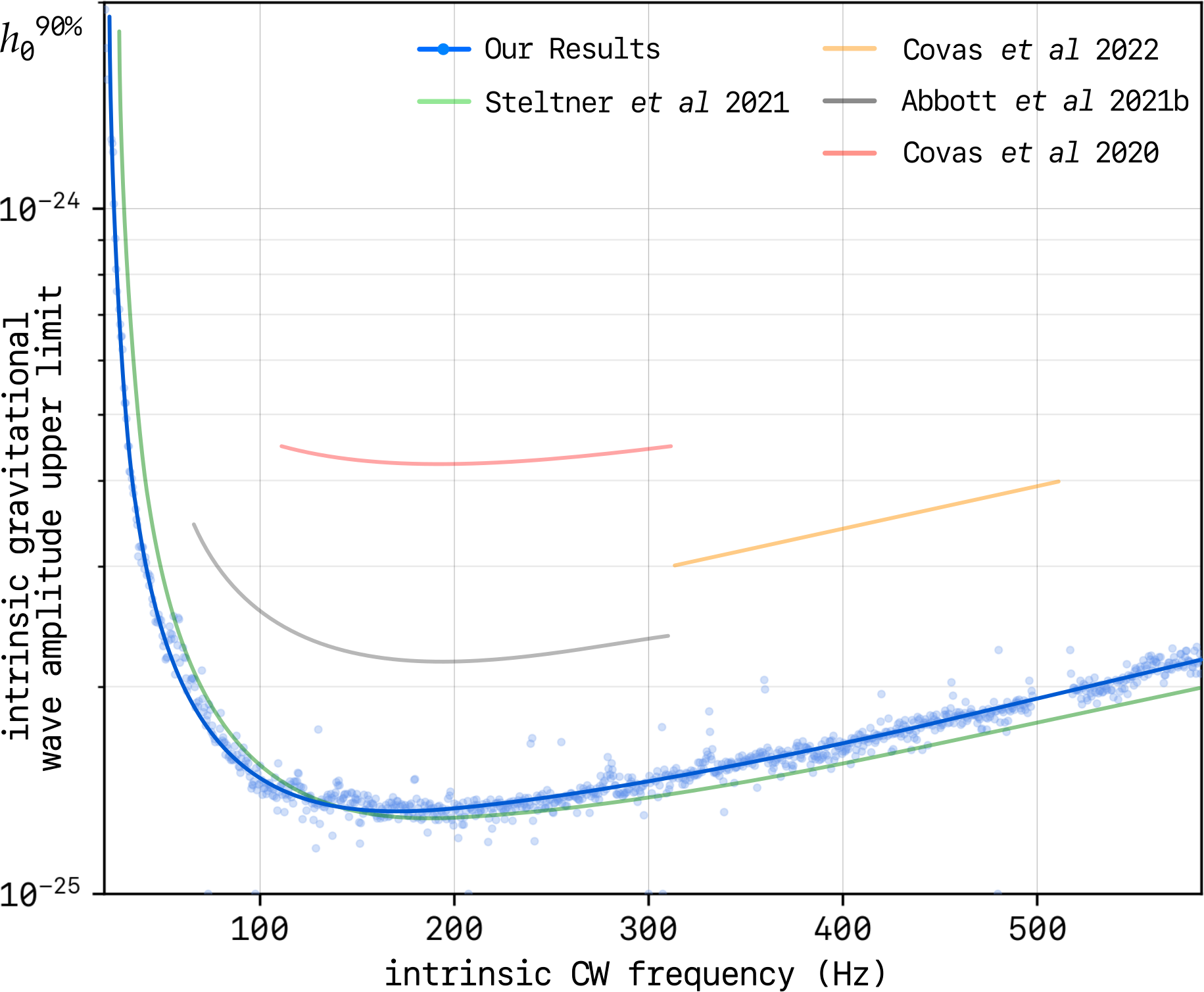}
\caption{{Constraints on the intrinsic amplitude of continuous gravitational waves ($\varsud{h}{90\%}{0}$) from neutron stars in binary orbits populating indistinguishable and distinguishable regions, as a function of the gravitational wave frequency (most likely twice the rotation frequency). To ease comparisons and guide the eye, we have plotted `smoothened' results from previous searches for binary pulsars in LIGO data \citep{O2BinPepSintes}, \citep{O2BinPepPapa} and \citep{O3BinLIGO}; \avn{these searches cover a different part of binary parameter space, where dedicated and costly searches are necessary. The parameter ranges are shown in Fig.~\ref{fig:reg}.}}}
\label{fig:ul}
\end{figure}

\begin{figure}[ht!]
\centering\includegraphics[width=85mm]{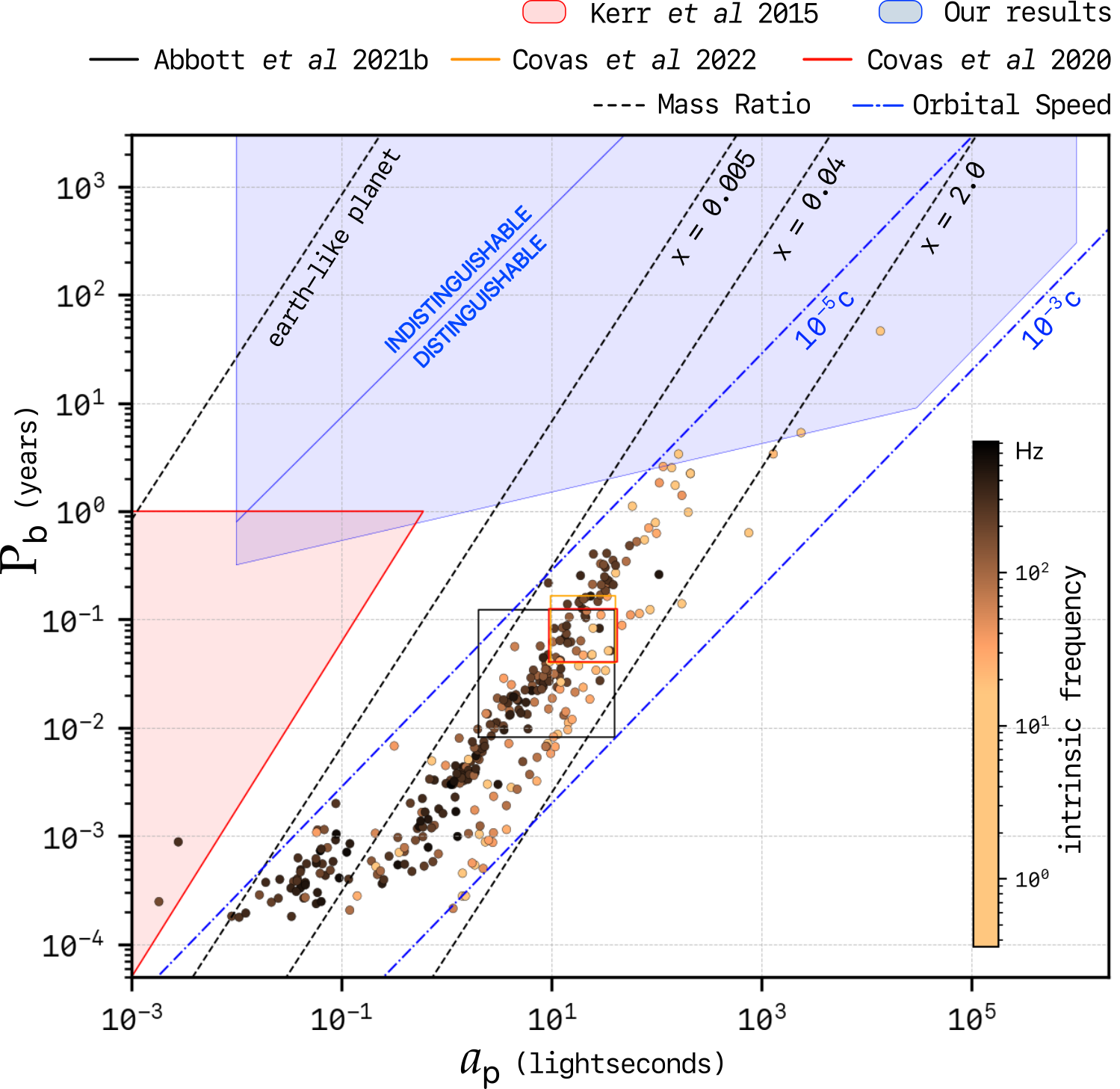}
\caption{{ATNF catalogue binary pulsars plotted in the \avn{$\vars{a}{p}-\vams{P}{b}$} plane and color-coded by rotational frequency. The blue lines mark the contours of constant orbital speed, while the black lines demarcate contours of constant mass ratios ($x=\vams{M}{c}/ 1.4 \sm$). The area where pulsar companions are deemed rare by \cite{Planets2015} is shaded in red; the areas constrained in this paper based on the results of \cite{Steltner:2020hfd}, are the ``distinguishable" and ``indistinguishable" areas above. The boxes mark the regions searched by previous gravitational wave searches.}}
\label{fig:reg}
\end{figure}

In order to determine $h_0^{90\%}$, we add fake signals to the real data and search for them with exactly the same procedure as used in the search \citep{Steltner:2020hfd}, including data cleaning and candidate clustering. \avn{The upper limits are established in 0.5 Hz bands, so a set of a 500 injection-and-recoveries -- 100 per $h_0$ value -- are performed in every 0.5 Hz band. The parameters of the simulated signals are drawn from the target signal population uniformly within the 0.5 Hz band, uniformly in the sky, log-uniformly in spin-down, and uniformly in orientation and polarization. The binary orbital parameters of the simulated signals are randomly sampled from distinguishable and indistinguishable regions combined as shown in top panel of Figure \ref{fig:SNRchangeDistr} (or blue region in Figure \ref{fig:reg})}. The resulting 90\% constraints are shown in Figure \ref{fig:ul}. The sensitivity of the isolated searches to binary orbits degrades slightly with increasing frequency due to the linear dependence of the binary modulation on frequency. This is reflected in the slight worsening of constraints in our results (blue) above roughly $200$ Hz in comparison to \citep{Steltner:2020hfd} (green).

\section{Conclusions}
\label{sec:conc}
We have demonstrated that simply based on search results for signals from isolated objects, the gravitational wave amplitude of signals from neutron stars in long period binary systems (upper blueish region in Figure \ref{fig:SNRchangeDistr}) can be constrained with remarkable sensitivity at practically zero cost. \avn{While these upper limits may be impressive considering the sheer volume of the binary parameter space considered, they are nonetheless set in a region which is computational cheaper to explore with implicit studies such as this one. In contrast, other recent searches \citep{O3BinLIGO,O2BinPepPapa,O2BinPepSintes} have explored binary parameter spaces %in combined ranges of $\vars{a}{p}\in[2,40]$ lightseconds and $\vams{P}{b}\in[3,60]$ days 
(see Figure \ref{fig:reg}), where implicit searches are useless, leaving computationally expensive explicit searches as the only option. This impacts both the `breadth' and the sensitivity of those searches.}

\avn{Even in the parameter space considered here, the actual detection} of a signal would require an additional search, but our results demonstrate that such search could be limited to following-up the candidates from the first stage of an isolated search. This allows to piggy-back on results obtained at a very large computational cost and opens up a broad region of parameter space for investigation at sensitivities close to the highest levels achievable by continuous wave surveys from isolated objects.

\avn{Previous searches for continuous gravitational waves have understandably focused on the $(\vars{a}{p}, \vams{P}{b})$ regions where pulsars in binaries have been observed. In contrast, the region that we have examined in this study is nearly void of binaries containing known pulsars (see Fig.~\ref{fig:reg}). Nevertheless, even if not commonly observed, pulsars with light companions in the orbital parameter range that we have investigated do exist. For instance, while there appear to be no {\it{binary}} systems noted in the ATNF pulsar catalogue with period longer than 30 days and a companion lighter than $0.056\,\sm$ ($x< 0.04$),} four {\it{3-body}} systems are suspected or known to contain a pulsar with a companion in our orbital parameter range -- PSR B1620-26 \citep{Ford:1999mh,Thorsett:1999nc}, PSR B1257+12 \citep{Wolszczan:1992zg}, and perhaps PSR B0943+10 \citep{Starovoit:2021,B0943-10,Shaw:2022mxv} and PSR J2007+3120 \citep{Nitu}. The first two are millisecond pulsars, the last two spin at about 1 Hz, outside of the high sensitivity band of the current generation of gravitational wave detectors. The PSR B1620-26 is a pulsar-white-dwarf-planet system, \avn{with a planet of mass $2.5\,\mathrm{M}_{\textrm{Jup}}$ ($x = 0.0017$)} in a 100 year orbit around the two stars at a distance of about $10^4$ lightseconds. The PSR B1257+12 system is composed of a pulsar and three planets, in orbits with \avn{$\vams{P}{b}\sim$} 25, 67, 98 days and $\vars{a}{p}\sim$ 94, 180, 230 lightseconds, respectively. The planets have masses $x = 4.3\times10^{-8},\,9.2\times10^{-6},\,8.4\times10^{-6}$, respectively. 
\par\noindent

\avn{Furthermore, the paucity of neutron stars with low mass companions in the investigated binary parameter space may not be completely representative of the true number of existing systems;} since many new pulsars are monitored for relatively short periods of time, timing noise and the correlation of the orbital modulation with the intrinsic spin-down make it more likely that binary pulsars with low-mass companions and/or long-periods may be misclassified as isolated, such as for PSR J1024-0719 \citep{Kaplan:2016ymq}. Lastly, there is some evidence that planetary mass companions are associated with low radio-luminosity pulsars \citep{Spiewak:2017mzc}, making these systems hard to detect. These considerations have motivated dedicated searches for pulsars with companions \citep{Antoniadis:2020gos,Nitu}, which however have not conclusively explored the entire binary parameter space that we consider here.

\avn{A continuous wave signal from a binary pulsar with a planetary companion will be very interesting since it will help shed light on how these systems are formed, which is still an open question. Post-supernova planet formation scenarios involve proto-planetary disks, and the open question relates to the origin of the disk -- whether it forms from the supernova ejecta or from a previous companion which got irradiated during the supernova, or from the merger and tidal disruption of a carbon/oxygen white dwarf by the neutron star} \citep[see][and references therein]{PulsarTimingAndExoplanets}. \avn{A broader sample of observed systems would likely enable a decision among the different scenarios, and a different probe -- such as gravitational waves -- will prove immensely useful in interpreting the observations with its completely different ``visibility" priors than electromagnetic (EM) observations. For instance, gravitational waves from a recycled neutron star with a planetary companion would further weaken the supernova fallback hypothesis. Continuous gravitational waves could also provide accurate timing information for high-sensitivity EM pulsation searches and thanks to a joint EM-GW observation, the gravitational wave emission mechanism could be identified, elucidating the physics of the source.}

\avn{The $(\vars{a}{p}, \vams{P}{b})$ region that we have investigated, also includes neutron stars with companions much heavier than a planet (top-right section of Fig.~\ref{fig:reg} to the right of the $x=0.04$ contour line), in wide-orbit binaries. These systems likely result from either dynamical formation pathways, or if the companion is massive enough to the supernova event and any resulting kicks. A number of such objects are known to exist, although typically in eccentric orbits. One of those systems, J1840-0643 \citep{Knispel:2013da}, has twice its spin frequency within the range of ground-based gravitational wave detectors at roughly $52.6$ Hz and has been timed accurately enough to allow for a targeted search \citep{LIGOScientific:2021hvc}. Targeted searches are by construction the most sensitive continuous wave searches up to ten times more sensitive than broadband surveys. Consequently, the upper limit from this search at that frequency is about a factor of ten less constraining than the value reported by \cite{LIGOScientific:2021hvc}. But there may yet be more systems like J1840-0643 which have gone undetected by EM searches so far, and which broadband gravitational wave searches in the parameter space considered here could instead reveal.} 
\section{Acknowledgments} 
\label{sec:ack}
We are very grateful to Michael Kramer and John Antoniadis for insightful discussions. We thank Benjamin Steltner for his support in implementing the same pipeline as the Stage 0 of \cite{Steltner:2020hfd}, necessary for the upper limits set by this search, and Bruce Allen for his useful comments. The simulations for this study were performed on the ATLAS cluster at MPI for Gravitational Physics/Leibniz University Hannover. 

\bibliography{Bibliography}
\bibliographystyle{aasjournal}

\end{document}

%% file: Macros.tex
\def\avn#1{{{\color{black}#1}}}
\def\EatH{Einstein@Home}
\def\SNR{{SNR}}

\def\csin{\text{\,\it{sin}\,}}

\def\F{\mathcal{F}}
\def\TwoF{2\mathcal{F}}

\def\B{{\hat{\beta}_{\mathrm{S/GLtL}}}}

\newcommand{\vars}[2]{#1_\mathrm{#2}}

\newcommand{\vams}[2]{\mathrm{#1}_\mathrm{#2}}

\newcommand{\varsud}[3]{#1^\mathrm{#2}_\mathrm{#3}}

\def\sm{\mathrm{M}_\odot}